\documentclass{article}
\usepackage{amsmath}
\usepackage{graphicx}
\usepackage{amsfonts}
\usepackage{amssymb}
\begin{document}

\title{Anatomy of extreme events in a \\complex adaptive system}
\author{Paul Jefferies$^{1}$, David Lamper$^{2}$ and Neil F. Johnson$^{1}$\\$^{1}$Physics Department, Oxford University, Oxford, OX1 3PU, U.K.\\$^{2}$Oxford Centre for Industrial and Applied Mathematics,\\Oxford University, Oxford, OX1 3LB, U.K.}
\date{\today}
\maketitle
\begin{abstract}
We provide an analytic, microscopic analysis of extreme events in an adaptive
population comprising competing agents (e.g. species, cells, traders,
data-packets). Such large changes tend to dictate the long-term dynamical
behaviour of many real-world systems in both the natural and social sciences.
Our results reveal a taxonomy of extreme events, and provide a microscopic
understanding as to their build-up and likely duration.
\end{abstract}

\noindent{PACS: 87.23.Ge, 02.50.Le, 05.45.Tp, 87.23.Kg}

\newpage

Large unexpected changes or `extreme events' (e.g. crashes in financial
markets, or punctuated equilibria in evolution) happen infrequently, yet tend
to dictate the long-term dynamical behaviour of real-world systems in
disciplines as diverse as biology and economics, through to ecology and
evolution. The ability to generate large internal, so-called \emph{endogenous}
changes is a defining characteristic of complex systems, and arguably of
Nature and Life itself~\cite{sornette99,johansen01,bak98}. Such changes are
manifestations of subtle, short-term temporal correlations resulting from
internal collective behaviour. They seem to appear out of nowhere and have
long-lasting consequences. To what extent can they ever be `understood'?
Followers of the self-organized criticality view~ \cite{bak98} would claim
this question is na\"{i}ve because of an inherent self-similarity in Nature:
any large changes are simply magnified versions of smaller changes, which are
in turn magnified versions of even smaller changes, and so on. Such
self-similarity is presumed to underlie the power-law scaling observed in
natural, social and economic phenomena~\cite{bak98}. However, there are
reasons for believing that the largest changes may be `special' in a
microscopic sense~ \cite{johansen01}. Power-law scaling is only approximately
true, and does not apply over an infinite range of scales. Apart from being
atomistic at the smallest scale, a population of competing agents cannot cause
any effect larger than the population size itself: in short, the largest
changes will tend to `scrape the barrel' in some way.
Reference~\cite{johansen01} quotes Bacon from Novum Organum: ``Whoever knows
the ways of Nature will more easily notice her deviations; and, on the other
hand, whoever knows her deviations will more accurately describe her ways''.

This paper addresses the task of understanding, and eventually controlling,
the large endogenous changes arising in a complex adaptive system comprising
competing agents (e.g.\ species, cells, traders, data-packets). Our work
reveals a taxonomy of large changes, and provides a quantitative microscopic
description of their build-up and duration. Our results also provide insight
into how a `complex systems manager' might contain or control such extreme events.

We consider a generic complex system in which a population of
$N_{tot}$ heterogeneous agents with limited capabilities and
information, repeatedly compete for a limited global resource. Our
model was introduced in Ref.~ \cite{johnson00}, and is a
generalization of the El Farol bar problem and the Minority game,
concerning a population of people deciding whether to attend a popular
bar with limited seating~\cite{arthur94}. At timestep $t$, each agent
(e.g.\ a bar customer, or a market trader) decides whether to enter a
game where the choices are option 1 (e.g.\ attend the bar, or buy) and
option 0 (e.g.\ go home, or sell): $N_{0}$ agents choose 0 while
$N_{1}$ choose 1. The `excess demand' $D[t]=N_{1}-N_{0}$ (which mimics
price-change in a market) and number $V[t]=N_{1}+N_{0}$ of active
agents (which mimics volume of market orders) represent output
variables. These two quantities fluctuate with time, and can be
combined to construct other global quantities of interest for the
complex system studied (e.g.\ summing the price-changes gives the
current price). This model can reproduce statistical and dynamical
features similar to those in a
real-world complex adaptive system, namely a financial market~\cite{johnson00}%
, and exhibits the crucial feature of seemingly spontaneous large changes of
variable duration~\cite{johnson00,lamper02}. The resulting time-series appears
`random' yet is non-Markovian, with subtle temporal correlations which put it
beyond any random-walk based description. The temporal correlations of
price-changes and volume, and their cross-correlation, are of intense interest
in financial markets where so-called chartists offer a wide range of
rules-of-thumb~\cite{blair96} such as `volume goes with price trend'. Although
such rules are unreliable, the intriguing question remains as to whether there
could \emph{in principle} be a `science of charting'.

A subset $V[t]\leq N_{tot}$ of the population, who are sufficiently confident
of winning, are active at each timestep. For $N_{1}<N_{0}$ the winning
decision is 1 and vice-versa, i.e.\ the winning decision is given by
$H[-D[t]]$ where $H[x]$ is the Heaviside function. The global resource level
is so limited, or equivalently the game is so competitive, that at least half
the active population lose at each timestep~\cite{arthur94}. The only global
information available to the agents is a common bit-string `memory' of the $m$
most recent outcomes. Consider $m=2$; the $P=2^{m}=4$ possible history
bit-strings are 00, 01, 10 and 11, which can also be represented in decimal
form: $\mu\in\{0,1,\ldots,P-1\}$. A strategy consists of a response, $a^{\mu
}\in\{-1,1\}$ to each possible bit-string $\mu$, $a^{\mu}=1\Rightarrow$ option
1, and $a^{\mu}=-1\Rightarrow$ option 0. Hence there are $2^{P}=16$ possible
strategies. The heterogeneous agents randomly pick $s$ strategies each at the
outset, and update the scores of their strategies after each timestep with the
reward function $\chi\lbrack D]=\operatorname*{sgn}[-D]$, i.e.\ $+1$ for
choosing the minority action, $-1$ for choosing the majority action. Agents
have a time horizon $T$ over which strategy points are collected, and a
threshold level $r$ which mimics a `confidence'. Only strategies having $\geq
r$ points are used, with agents playing their highest scoring strategy. Agents
with no such strategy become temporarily inactive~\cite{johnson00}. We focus
on the regime where the number of strategies in play is comparable to the
total number available, since this yields seemingly random dynamics with
occasional large movements~ \cite{johnson00,lamper02}. The coin-tosses used to
resolve ties in decisions (i.e.\ $N_{0}=N_{1}$) and active-strategy scores,
inject stochasticity into the game's evolution. Reference~\cite{johnson02}
showed that a simplified version of this system in the limit $r\rightarrow
-\infty$ and $T\rightarrow\infty$, can be usefully described as a
stochastically disturbed deterministic system. We are interested in the
dynamics of large changes, and adopt the approach and terminology of
Ref.~\cite{johnson02}. Averaging over our model's stochasticity yields a
description of the game's deterministic dynamics via mapping equations for the
strategy score vector $\underline{S}[t]$ and global information $\mu\lbrack
t]$. For $s=2$ the deterministic dynamics are given exactly by the following
equations:
\begin{align}
\underline{S}[t]  &  =\underline{S}[0]-\sum_{i=t-T}^{t-1}\underline{a}%
^{\mu\lbrack i]}\operatorname*{sgn}\left[  D[i]\right]  ,\label{eq:sscore}\\
\mu\lbrack t]  &  =2\mu\lbrack t-1]-PH\left[  \mu\lbrack t-1]-P/2\right]
+H\left[  D[t-1]\right]  .\nonumber
\end{align}
The corresponding demand function is given by
\[
D[t]=\sum_{R=1}^{2P}a_{R}^{\mu\lbrack t]}H[S_{R}-r]\sum_{R^{\prime}=1}%
^{2P}\left(  1+\operatorname*{sgn}\left[  S_{R}[t]-S_{R^{\prime}}[t]\right]
\right)  \Psi_{R,R^{\prime}},
\]
where $\underline{\underline{\Psi}}$ is the symmetrized strategy allocation
matrix which constitutes the \emph{quenched disorder} present during the
system's evolution~\cite{johnson02}. Elements $\Psi_{R,R^{\prime}}$ enumerate
the number of agents holding both strategy $R$ and $R^{\prime}$. The volume
$V[t]$ is given by the same expression as $D[t]$ replacing $a_{R}^{\mu\lbrack
t]}$ by unity.

Large changes such as financial market crashes, seem to exhibit a wide range
of possible durations and magnitudes making them difficult to capture using
traditional statistical techniques based on one or two-point probability
distributions~\cite{johansen01}. A common feature, however, is an obvious
trend (i.e.\ to the eye) in one direction over a reasonably short time window:
we use this as a working definition of a large change. In fact, all the large
changes discussed here represent $>3\sigma$ events. In both our model and the
real-world system, these large changes arise more frequently than would be
expected from a random-walk model~\cite{sornette99,johansen01}. Our model's
dynamics can be described by trajectories on a de Bruijn graph~
\cite{johnson02}: see Fig.\ 1 for $m=3$, with a transition incurring an
increment to the score vector $\underline{S}$. There are $P$ orthogonal
increment vectors $\underline{a}^{\mu}$, one for each node $\mu$. Setting the
initial scores $\underline{S}[0]=\underline{0}$, the strategy score vector in
Eq.~(\ref{eq:sscore}) can be written exactly as:
\[
\underline{S}[t]=c_{0}\underline{a}^{0}+c_{1}\underline{a}^{1}+\ldots
+c_{P-1}\underline{a}^{P-1}\ =\ \sum_{j=0}^{P-1}c_{j}\underline{a}^{j}%
\]
where $c_{j}$ represents the \emph{nodal weights} for history node $\mu=j$.
The nodal weights enumerate the number of negative return transitions from
node $\mu$ minus the number of positive return transitions, in the time window
$t-T\rightarrow t-1$. High absolute nodal weight implies persistence in
transitions from that node i.e.\ persistence in $D|\mu$. Large changes will
occur when connected nodes become persistent. The simplest type of large
movement exhibiting perfect nodal persistence would be $\mu=0,0,0,0,\ldots$ in
which all successive price changes are in the \emph{same} direction. We call
this a `fixed-node crash' (or rally). However, there are many other
possibilities reflecting the wide range of forms and durations of the large
change. For example, on the $m=3$ de Bruijn graph in Fig. 1 the cycle
$\mu=0,0,1,2,4,0,\ldots$ has four out of the five transitions producing
price-changes of the same sign (it is persistent on nodes 1, 2, 4 and
antipersistent on node 0). We call this a `cyclic-node crash' (or rally).
Figure 2 illustrates a large change which starts as a fixed-node crash then
subsequently becomes a cyclic-node crash. Cyclic-node crashes can be treated
simply as interlocking fixed-node crashes, hence for clarity we focus here on
a single fixed-node crash (or rally). For the parameter ranges of interest,
the choice about whether a strategy is played by an agent is more determined
by whether that strategy's score is above the threshold, than whether it is
their highest-scoring strategy~\cite{supplement}. This is because agents are
only likely to have at most one strategy whose score lies above the threshold
for confidence levels $r\geq0$. Making the additional numerically-justified
approximation of small quenched disorder (i.e.\ the variance of the entries in
the strategy allocation matrix $\underline{\underline{\Psi}}$ is smaller than
their mean for the parameter range of interest~\cite{johnson02}), the demand
and volume become:
\begin{align}
D[t]  &  = \frac{N}{4P}\sum_{R=1}^{2P}a_{R}^{\mu\lbrack t]}%
\operatorname*{sgn}\left[  S_{R}[t]-r\right]  ,\label{eq:sqddemand}\\
V[t]  &  = \frac{N}{2}+\frac{N}{4P}\sum_{R=1}^{2P}\operatorname*{sgn}%
\left[  S_{R}[t]-r\right]  .\label{eq:sqdvolume}%
\end{align}

Suppose persistence on node $\mu=0$ starts at time $t_{0}$. How long will the
resulting crash last? To answer this, we decompose Eq.~(\ref{eq:sqddemand})
into strategies which predict 1 at $\mu=0$, and those that predict 0. We first
consider the particular case where the node $\mu=0$ was \emph{not} visited
during the previous $T$ timesteps, hence the loss of score increment from
time-step $t-T$ will not affect $\underline{S}[t]$ on average. At any later
time $t_{0}+\tau$ during the crash, (i.e.\ $\mu=0$) Eqs.~(\ref{eq:sqddemand})
and (\ref{eq:sqdvolume}) are hence given by:
\begin{align}
D[t_{0}+\tau] &  = -\frac{N}{4P}\left\{  \sum_{R\ni a_{R}^{\mu}%
=-1}\operatorname*{sgn}\left[  S_{R}[t_{0}]-r-\tau\right]  -\sum_{R\ni
a_{R}^{\mu}=1}\operatorname*{sgn}\left[  S_{R}[t_{0}]-r+\tau\right]  \right\}
,\label{eq:sqddemand2}\\
V[t_{0}+\tau] &  = \frac{N}{2}+\frac{N}{4P}\left\{  \sum_{R\ni a_{R}%
^{\mu}=-1}\operatorname*{sgn}\left[  S_{R}[t_{0}]-r-\tau\right]  +\sum_{R\ni
a_{R}^{\mu}=1}\operatorname*{sgn}\left[  S_{R}[t_{0}]-r+\tau\right]  \right\}
.\nonumber
\end{align}
$|D[t_{0}+\tau]|$ decreases as the persistence time $\tau$ increases, and
hence the crash ends at time $t_{0}+\tau_{c}$ when the right-hand side of
Eq.~(\ref{eq:sqddemand2}) changes sign. The persistence time or `crash-length'
$\tau_{c}$ is thus given by the mean of the scores of the strategies
predicting 0, i.e.\ $\tau_{c}=\overline{S}_{R\ni a_{R}^{\mu}=-1}[t_{0}%
]=-c_{0}[t_{0}]$. In the more general case where the node $\mu=0$ \emph{was}
visited during the previous $T$ timesteps, $\tau_{c}$ is given by the largest
$\tau$ value which satisfies:
\[
\tau=-\left(  c_{0}[t_{0}]+\sum_{\{t^{\prime}\}}\operatorname*{sgn}\left[
D[t^{\prime}]\right]  \right)
\]
where $\{t^{\prime}\}\ni(\mu\lbrack t^{\prime}]=0\cap t_{0}-T\leq t^{\prime
}\leq t_{0}+\tau-T)$. Assume that the scores have a near-Normal
distribution, i.e.\ $S_{R\ni a_{R}^{\mu}=-1}[t_{0}]\sim\operatorname*{N}%
[\overline{S}_{-1},\sigma]$ as in Fig.~3a. For each strategy $R$ there exists
an anticorrelated strategy $\overline{R}$ and hence $S_{R}[t]=-S_{\overline
{R}}[t]$ for all $t$. Consequently, prior to a crash, the score distribution
tends to split into two halves as indicated schematically in Fig. 3a. The
expected demand (and volume) during the crash are then:
\begin{align*}
<D[t_{0}+\tau]> &  \propto\left(  \operatorname{erf}\left[  \frac{c_{0}%
[t_{0}]+r+\tau}{\sqrt{2}\sigma}\right]  -\operatorname{erf}\left[
\frac{-c_{0}[t_{0}]+r-\tau}{\sqrt{2}\sigma}\right]  \right) \\
<V[t_{0}+\tau]> &  \propto\left(  2-\operatorname{erf}\left[  \frac
{c_{0}[t_{0}]+r+\tau}{\sqrt{2}\sigma}\right]  -\operatorname{erf}\left[
\frac{-c_{0}[t_{0}]+r-\tau}{\sqrt{2}\sigma}\right]  \right)
\end{align*}
These forms are illustrated in Figure 3b. As the spread in the strategy score
distribution is increased, the dependence of $<D>$ and $<V>$ on the parameters
$\tau$ and $r$ becomes weaker and the surfaces flatten out leading to a
smoother drawdown, as opposed to a sudden severe crash. As the parameters
$\overline{S}_{-1},\sigma,r$ are varied, it can be seen that the behaviour of
the demand and volume during the crash can exhibit markedly different
qualitative forms~\cite{supplement}, yielding a \emph{taxonomy} of
\emph{different species} of large change \emph{even within the same
single-node family}. This result could explain why financial market chartists'
rules-of-thumb~\cite{blair96}, such as `volume goes with price trend', are far
too simplistic.

We now turn to the important practical question of whether history will repeat
itself, i.e.\ given that a crash has recently happened, is it likely to happen
again? If so, is it likely to be even bigger? Suppose the system has built up
a negative nodal weight for $\mu=0$ at some point in the game (see Fig.\ 4a).
It then hits node $\mu=0$ at time $t_{0}$ producing a crash (Fig.\ 4b). The
nodal weight $c_{0}$ is hence restored to zero (Fig.\ 4c). In this model the
previous build-up is then forgotten because of the finite $T$ score window,
hence $c_{0}$ becomes positive (Fig.\ 4d). The system then corrects this
imbalance (Fig.\ 4e), restoring $c_{0}$ to 0. The crash is then forgotten,
hence $c_{0}$ becomes negative (Fig.\ 4f). The system should therefore crash
again - however, a crash will \emph{only} re-appear if the system's trajectory
subsequently returns to node $\mu=0$. Interestingly, we find that the
\emph{disorder} in the initial distribution of strategies among agents
(i.e.\ the quenched disorder in $\underline{\underline{\Psi}}$) can play a
deciding role in the issue of crash `births and revivals' since it leads to a
slight bias in the outcome, and hence the subsequent transition, at each node.
When $c_{\mu\lbrack t]}=0$ (see Fig.\ 4c), it follows that
$\operatorname*{sgn}[D[t]]$ is more likely to be equal to $\operatorname*{sgn}%
\left[  \underline{a}^{\mu\left[  t\right]  }\cdot\underline{x}\right]  $
where $\underline{x}=\sum_{R^{\prime}=1}^{2P}\underline{\underline{\Psi}%
}_{R^{\prime}}$ is a strategy weight vector with $x_{R}$ corresponding to the
number of agents who hold strategy $R$~\cite{supplement}. The quenched
disorder therefore provides a crucial bias for determining the future
trajectory on the de Bruijn graph when the nodal weight is small, and hence
can decide whether a given crash recurs or simply disappears. The quenched
disorder also provides a \emph{catalyst} for building up a very large crash.

Our work opens up the study of how a `complex-systems-manager' might use this
information to control the long-term evolution of a complex system by
introducing, or manipulating, such large changes. As an example, we give a
quick three-step solution to prevent large changes: (1) use the past history
of outcomes to build up an estimate of the score vector $\underline{S}[t]$ and
the nodal weights $\{c_{\mu\lbrack t]}\}$ on the various critical nodes, such
as $\mu=0$ in the case of the fixed-node crash. (2) Monitor these weights to
check for any large build-up. (3) If such a build-up occurs, step in to
prevent the system hitting that node until the weights have decreased.

%\bigskip
%

\newpage

\textbf{Figure Captions}

\bigskip

Figure 1: Dynamical behaviour of the global information is described by
transitions on the de Bruijn graph. Graph for population of $m=3$\ agents.
Blue transitions represent positive demand $D$, red transitions represent
negative demand. \bigskip

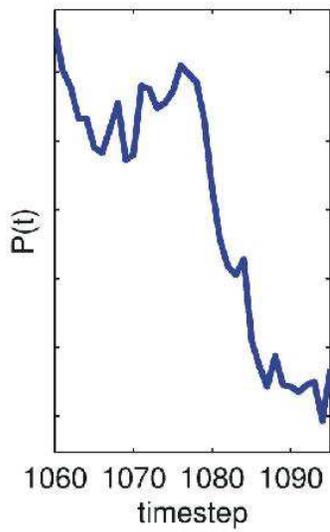
Figure 2: Dynamical behaviour of complex system (e.g. price $P\left[
t\right]  $\ in financial market) described by evolution of nodal weights
$c_{\mu}$. History at each timestep indicated by black square. Large change
preceded by abnormally high nodal weight. Large change incorporates fixed-node
and cyclic node crashes \bigskip

Figure 3: (a) Schematic representation of strategy score distribution prior to
crash. Arrows indicate subsequent motion during crash period. (b) Plots of
expected demand and volume during crash period showing range of different
possible behaviour as system parameters are varied. \bigskip

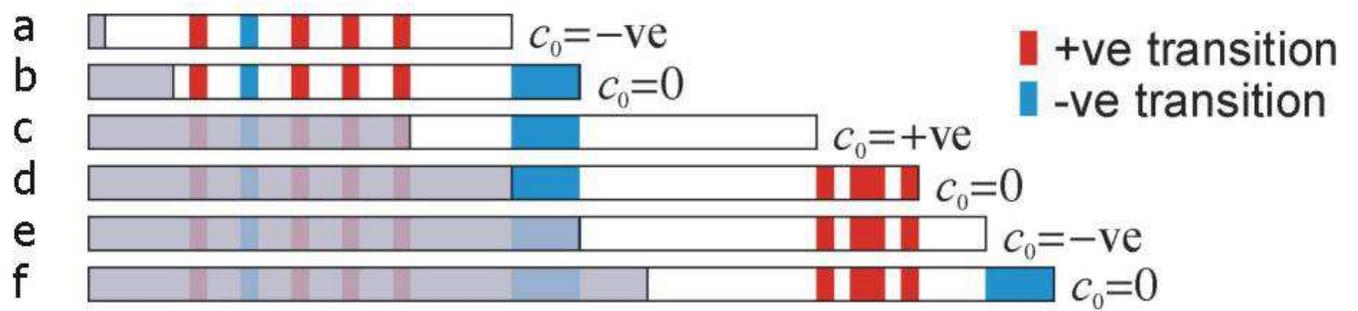
Figure 4: Representation of how large changes can recur due to finite memory
of agents. Grey area shows history period outside agents' memory. Example
shows recurring fixed-node crash at node $\mu=0$.
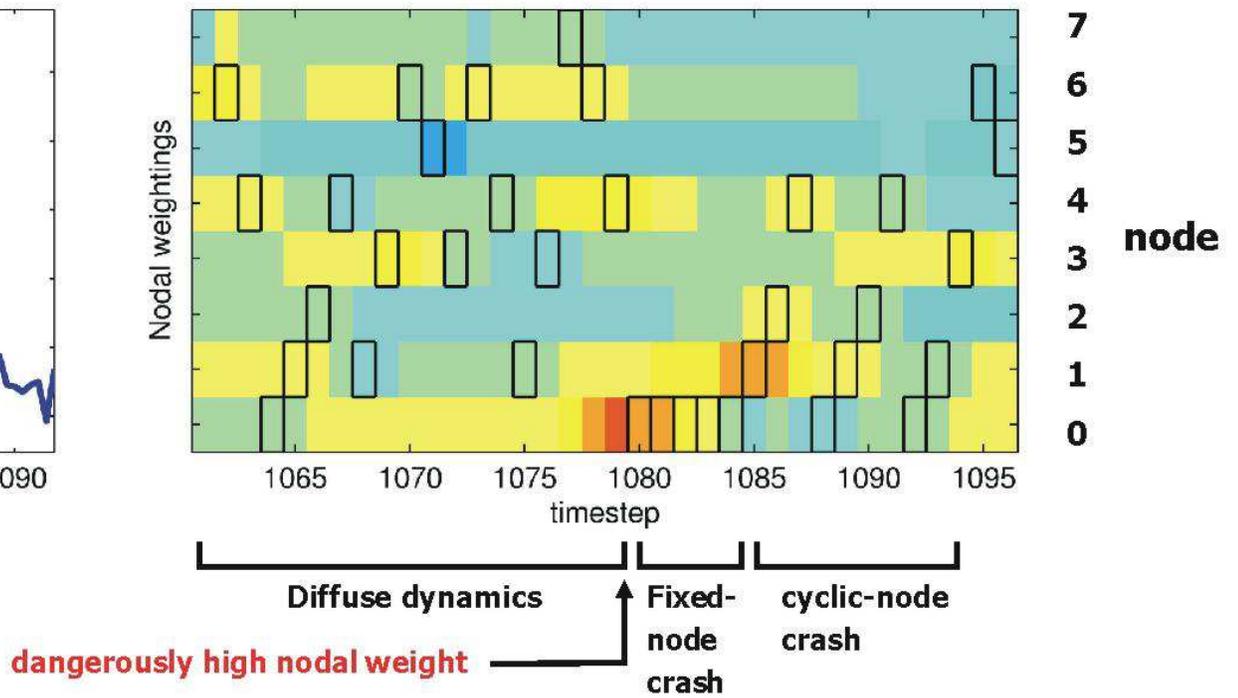
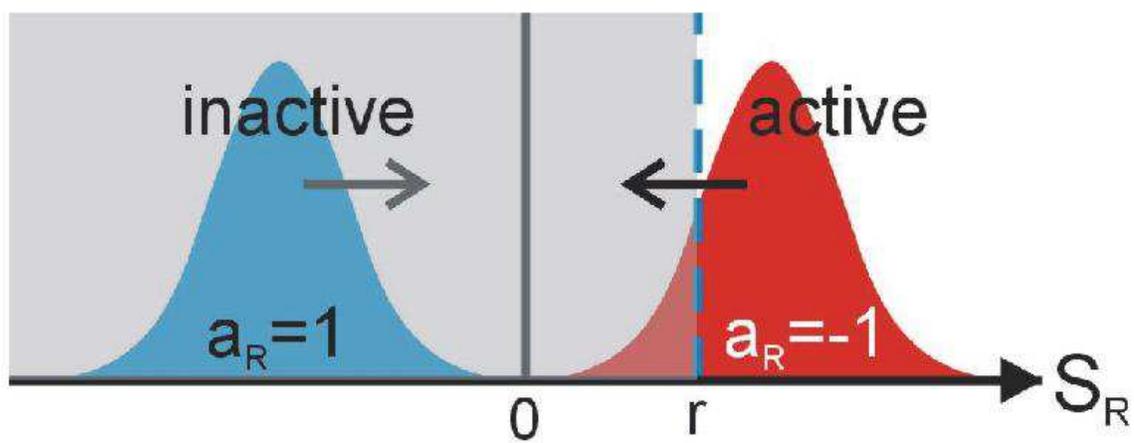
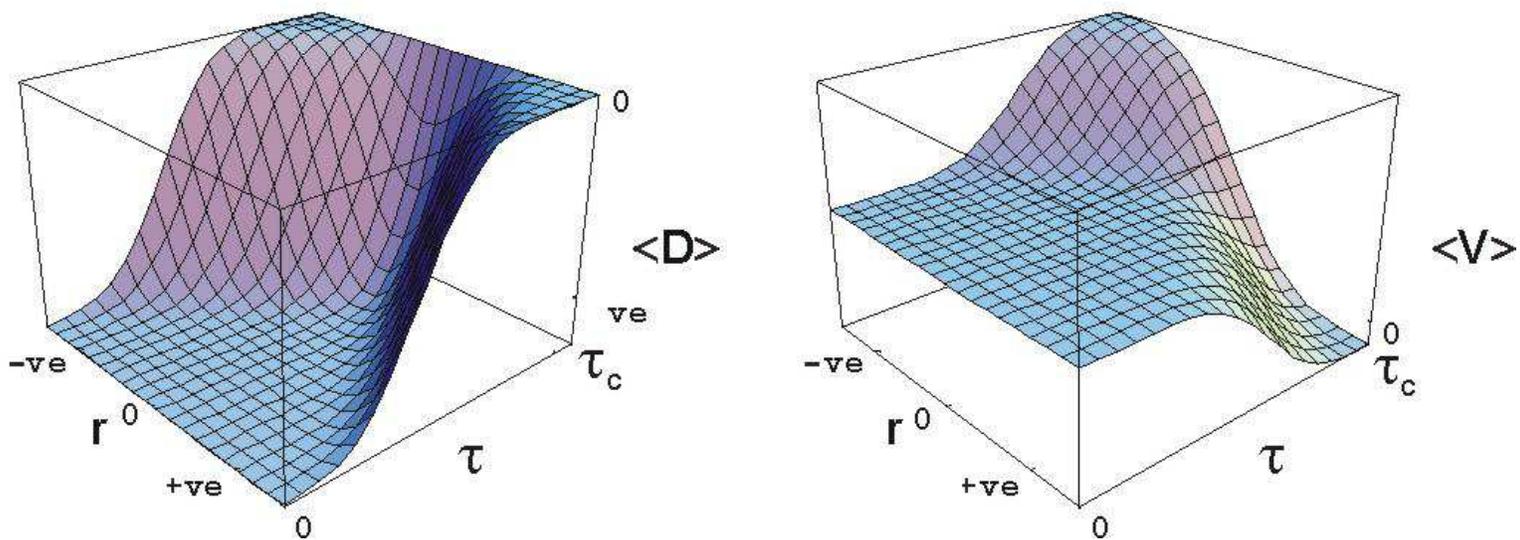
\end{document}